\documentclass[twocolumn,aps,prl,showpacs]{revtex4}
\usepackage{graphicx}
\begin{document}

\title{Pressure dependence of diffusion in simple glasses and supercooled 
liquids}
\author{H.~R.~Schober}
\affiliation{Institut f\"ur Festk\"orperforschung, Forschungszentrum
J\"ulich, D-52425, Germany}

\date{\today}
\begin{abstract}
Using molecular dynamics simulation, we have calculated the pressure 
dependence of the diffusion constant in a binary Lennard-Jones Glass.
We observe four temperature regimes. The apparent activation volume
drops from high values in the hot liquid to a plateau value. Near
the critical temperature of the mode coupling theory it rises steeply, but
in the glassy state we find again small values, similar to the ones in 
the liquid.  The peak of the activation volume at the critical temperature
is in agreement with the prediction of mode
coupling theory. 
 
\end{abstract}
\pacs{61.43.Fs,66.10.-x,66.30.-h,64.70.Pf}
\maketitle


Diffusion  in glasses and their melts has been studied intensively
for many years. These efforts are stimulated both by the technological 
importance of glassy and amorphous materials and by the desire to
understand the physics of disordered systems in general and the liquid
to glass transition in particular.
Despite this effort there is still no agreement on the
nature of diffusion on an atomic level or on its change at temperatures near
the glass transition.
This even holds for simple densely packed glasses, such as 
binary metallic ones, see for recent reviews 
Refs.~\onlinecite{mehrer:96,frank:97,loirat:00,faupel:01,RMP}. 

In a hot liquid, diffusion is by flow, whereas, in the
glass well below the transition temperature, it will be mediated by hopping
processes. One  key question is the transition between the
two regimes. For fragile glasses, such as most polymers and amorphous
metallic glasses, mode coupling theory (MCT) predicts an arrest
of the homogeneous viscous flow in the 
undercooled melt at a temperature $T_c$, well above the glass
transition temperature $T_g$. \cite{gotze:92} Hopping processes
will suppress the predicted singularities and will become the dominant
diffusion process near $T_c$.
 
The nature of the hopping process is another issue of controversy.
Is it by a vacancy mechanism, similar to diffusion in the crystalline state, 
or is it via a collective process inherent to the disordered
structure ? Investigations are hampered by the fact that
glasses are thermodynamically not in  equilibrium, and
one observes aging. The diffusion constant of a glass which has been
relaxed for a long time will be considerably lower than the one of an
``as quenched'' glass. 

In crystalline materials the pressure dependence of the diffusion constant
can often be used to identify the diffusion mechanism. For thermally
activated diffusion the diffusion constant can be described by an 
Arrhenius law
\begin{equation}
D(T) = D_0 \exp{\left( -H/kT \right)}
\end{equation}
where $D_0$ is a pre-exponential factor and $H$ is the activation enthalpy.
In a more component system the diffusion constants and the parameters
describing them will be different for each component. For simplicity's
sake we drop the indices indicating the component here and in the 
following, whenever possible.
 
Using $V=\partial G / \partial p$ with $G=H-TS$ one obtains the activation
volume for a diffusion by a single jump process \cite{mehrer:90}
\begin{equation}
V_{\rm act} = -kT \left[ \frac{\partial \ln{D}}{\partial p}\right]_T +
               kT \left[ \frac{\partial \ln{D_0}}{\partial p}\right]_T. 
\label{eq_v_cryst}
\end{equation}
In crystals one finds that the second term is only a minor correction
which can be neglected. For diffusion mediated by defects, the activation
volume splits into two terms, 
$V_{\rm act} = V^f + V^m$, a formation volume $V^f$ and a migration
volume $V^m$  

For diffusion via thermal vacancies the formation volume dominates and 
$V_{\rm act}$ varies between $0.6\Omega$ and $1 \Omega$, where $\Omega$
is the average atomic volume. For the migration part one estimates
$V^m \sim 0.1 \Omega$. Concomitantly in crystals high values of
$V_{\rm act}$ are taken as a signature of a thermally activated diffusion
mechanism. 

Assuming that also in the glass the first term in Eq.~\ref{eq_v_cryst}
dominates, one usually describes, also in amorphous materials, 
the pressure dependence of diffusion 
by an apparent activation volume 
\begin{equation}
\hat{V}_{\rm act} = -kT \left[ \frac{\partial \ln{D}}{\partial p}\right]_T
\label{eq_v}.
\end{equation}

Experiments on a number of metallic glasses give a large spread of 
values in the range of
0.05 to 1 $\Omega$.\cite{RMP} 
Low values were, e. g., observed in Co$_{81}$Zr$_{19}$ \cite{klugkist:98}
where no significant isotope effect is observed \cite{heesemann:95}.
This result can be interpreted in terms of a collective diffusion
mechanism inherent to the glassy structure.
The situation
is not so clear for the case of large activation volumes $\hat{V}_{\rm act}$.
The activation volume of $0.9$~$\Omega$ for diffusion of Zr in Co$_92$Zr$_8$
was interpreted as indication of diffusion via vacancy like defects
\cite{klugkist:98} rather than by an inherent mechanism as in the case Co.
Values of around $0.5 \Omega$ have been observed in several materials
\cite{hofler:92,grandjean:97,klugkist:99}. Such values are also found in
materials where the vanishing isotope points to diffusion by collective
jumps.\cite{heesemann:00} Whether collectivity can induce migrational
activation volumes of the order 0.5 to 1 $\Omega$ is still open.

In the liquid state the diffusion constant can be fitted by a 
Vogel-Fulcher-Tammann law  (VFT) 
\begin{equation}
D^{\rm VFT}(T) = D_0^{\rm VFT}\exp{\left[-E^{\rm VFT}/(T-T^{\rm VFT})\right]}.
\label{eq_VFT}
\end{equation}
From this we get the activation volume
\begin{eqnarray}
\hat{V}_{\rm act}^{\rm VFT} = && -kT \left[ 
\frac{\partial \ln{D_0^{\rm VFT}}}{\partial p} - 
\frac{1}{k(T-T^{\rm VFT})}\frac{\partial E^{\rm VFT} }{\partial p} 
 \right. \nonumber \\ 
&&-\left.
\frac{E^{\rm VFT}}{k^2(T-T^{\rm VFT})^2}\frac{\partial T^{\rm VFT} }{\partial p}\right]_T .
\label{eq_vVFT}
\end{eqnarray}
From MCT one derives \cite{gotze:92} alternative expressions
\begin{equation}
D^{\rm MCT}(T) = D_0^{\rm MCT}\left(T - T_c^{\rm MCT}\right)^\gamma 
\label{eq_MCT}
\end{equation}
\begin{eqnarray}
\hat{V}_{\rm act}^{\rm MCT} = && -kT \left[ 
\frac{\partial \ln{D_0^{\rm MCT}}}{\partial p} - 
\gamma \frac{1}{(T - T_c^{\rm MCT})}\frac{\partial T_c^{\rm MCT} }{\partial p}
\right. \nonumber \\
&& + \left. \ln{(T - T_c^{\rm MCT})}\frac{\partial \gamma }{\partial p}
\right]_T .
\label{eq_vMCT}
\end{eqnarray}
In both expressions the diffusion constant extrapolates to zero, however,
with one crucial difference. In the VFT-expression, Eq.~\ref{eq_VFT},
this happens at $T_0^{\rm VFT}$, far below $T_g$, whereas in MCT diffusion
vanishes at,  $T_c^{\rm MCT}$, well above $T_g$. Due to the crossover
to the glass the difference between the two expressions is 
normally not sufficient
to distinguish between them with certainty. This is different for the
pressure dependence where both  expressions give singularities, below
and above $T_g$, respectively.   

There is only one measurement of $\hat{V}_{\rm act}$ in the undercooled 
melt. Knorr {\it et al.}\cite{knorr:00} 
find for Ni diffusion in the bulk
metallic glass melt
Zr$_{46.75}$Ti$_{8.25}$Cu$_{7.5}$Ni$_{10}$Be$_{27.5}$ values
between 0.35 and 0.64~$\Omega$. 
  
Whereas experiment so far only provides circumstantial evidence for
the diffusion mechanism, molecular dynamics simulation can give both
the quantities seen in experiment and show the underlying atomic processes. 
In a previous simulation on NiZr the activation volume was estimated
as $0.36$~$\Omega$ from the change of barrier hight.\cite{teichler:01} 
The underlying
process was a collective jump of a chain of atoms.

The aim of the present paper is to present a systematic study of the 
pressure dependence of diffusion as function of temperature. In order
to relate closely to other work the simulations were done for the
well studied binary Lennard-Jones system (LJ), described by the inter-atomic
interaction:
\begin{equation}
V_{ij}(R) = 4 \epsilon_{ij} \left[ \left( \sigma_{ij} /R \right)^{12} - 
                         \left( \sigma_{ij} /R \right)^{6} +
                          A_{ij} R + B_ij \right] .
\label{eq_LJ}
\end{equation}
For the parameters $\epsilon_{ij}$ and $\sigma_{ij}$ we took the commonly
used values of Kob and Andersen \cite{kob:95a}: 
$\epsilon_{AA}= \epsilon = \sigma_{AA} =\sigma = 1$, 
$\epsilon_{BB}=0.5$, $\sigma_{BB}=0.88$,
$\epsilon_{AB}=1.5$ and $\sigma_{AB}=0.9$.  
Different from these authors we use a larger cutoff radius $R_c = 3\sigma$.
To avoid spurious cutoff effects we introduce,
similar to 
the shifted force potential \cite{nicolas:79}, the parameters
$A_{ij}$ and $B_{ij}$
to ensure continuity of the potential and its first derivative at the
cutoff. All masses were set to $m_j=1$.
In the following,  we will give all results in the 
reduced
units of energy, $\epsilon_{AA}$, length, $\sigma_{AA}$, and atomic mass 
$m_A$.

We have used this potential previously to calculate the diffusional
isotope effect parameter in the liquid \cite{S:01} and the evolution
of the dynamic heterogeneity both below and above the glass transition
\cite{CMS:00}. From these calculations we had well aged samples at
different temperatures for zero external pressure.

The calculations were done at each temperature with constant  
volume, corresponding to zero pressure, and periodic boundary 
conditions. The pressure was monitored. 
The time step was 
$\Delta t = 0.005$. A heat bath was simulated by comparing
the temperature, averaged over  20 time steps, with the nominal temperature.
At each  time step 1\% of the temperature difference was adjusted by
random additions to the particle velocities. This procedure assured that
existing correlations between the motion of atoms were only minimally
affected. At each temperature we had 8 independent 
samples, each of 5488 atoms in a ratio $4:1$ of $A$- and 
$B$-atoms. The diffusion constants were calculated from the asymptotic
slope of the atomic mean square displacements
\begin{equation}
D(T) = \frac{1}{6t} 
\lim_{t \to \infty}\langle ({\bf R^\ell}(t+t')-{\bf R^\ell}(t')
\rangle_{\ell,t'}   
\end{equation}
where $\langle  ..... \rangle_{\ell,t'}$ indicates averaging over all
atoms of the particular species and over starting times.
\begin{figure}
\includegraphics*[bb=30 100 530 500,totalheight=6cm,keepaspectratio]{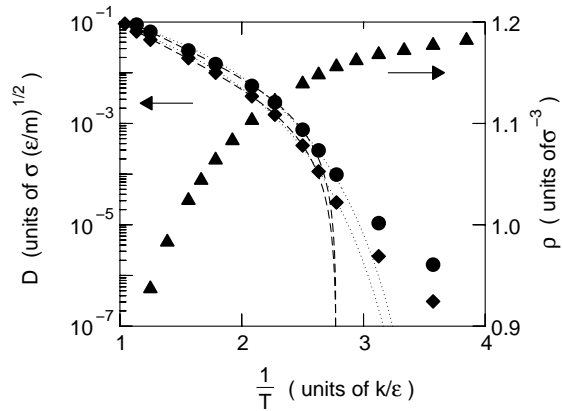}
\caption{Diffusion constants (majority $A$-atoms, diamonds, and minority 
$B$-atoms, 
circles) and density (triangles)
at zero pressure against inverse temperature (all in reduced units). 
The dashed and dotted lines
show the fits with the MCT and VFT expressions, respectively}
\label{fig_D_rho}
\end{figure}

Fig.~\ref{fig_D_rho} shows the densities and diffusion constants for the
zero pressure configurations. From the change in slope 
of the volume expansion
we estimate the glass transition temperature as $T_g \approx 0.35 \epsilon/k$.
The diffusion constant can be fitted very well both by the VFT and MCT
expressions. For both fits we assume that the respective transition 
temperatures are the same for both species. 
For the VFT expression, Eq.~\ref{eq_VFT}, we find the parameters
$T_0^{\rm VFT} = 0.244\epsilon/k$ for both species and
$E^{\rm VFT}= 1.092$ and $0.986 \epsilon$ for A and B atoms, respectively.
The values for the MCT-expression, Eq.~\ref{eq_MCT}, are
$T_c = 0.36 \epsilon /k$ for both species and 
$\gamma = 1.87$ and $\gamma = 2.02$ for A and B, respectively. 
The two temperatures,  $T_g$  and  $T_c$ are
very close to each other, but are much lower than the 
value $T_c = 0.435 \epsilon /k$, reported for simulations at constant 
density $\rho = 1.2$.\cite{kob:95b} This reflects the strong dependence
of the glass transition 
on density or pressure. We find for zero pressure a density of $\rho = 1.15$ 
at $T_g$. 

To calculate the pressure dependence we did additional runs with higher
and lower densities, respectively. At each temperature the zero pressure
samples were taken as starting configurations which were then compressed
(expanded) and subsequently aged. The change of density was 2\% at the
highest and 0.5\% at the lowest temperature. The resulting pressure
change after aging was kept this way within the approximate limits
of $\pm 0.2 \epsilon / \sigma^3$. Aging varied between 100 000 and
several million time steps, depending on temperature.

\begin{figure}
\includegraphics[bb=50 100 530 530,totalheight=6cm,keepaspectratio]{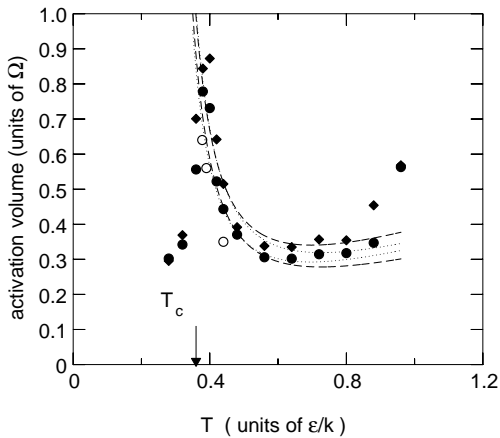}
\caption{Calculated activation volume (majority $A$-atoms, diamonds, 
and minority $B$-atoms, circles) versus temperature. The open circles
represent the experimental data on Ni diffusion in a bulk glass melt,
scaled to LJ-units. The dashed and dotted lines are fits with the
VFT expression, with one or two VFT temperatures, respectively.} 
\label{fig_vVFT}
\end{figure}

After calculating the diffusion constants the activation volume was
evaluated from Eq.~\ref{eq_v}. Fig.~\ref{fig_vVFT} shows the
resulting values together with the fit with the VFT-expression, 
Eq.~\ref{eq_vVFT}. One can clearly distinguish four temperature
regions. At the highest temperature 
$\hat{V}_{\rm act} \approx 0.6$~$\Omega$, nearly identical for 
both components. 
Upon cooling  the activation
volume drops to a plateau value $\hat{V}_{\rm act} \approx 0.3$~$\Omega$.
The larger component has, as one would expect, a slightly higher
activation volume.  Below $T=0.5$ the activation volume rises sharply
for both components and reaches a maximum of order
$\Omega$ around $T=0.4$, near $T_g$ and 
$T_c^{\rm MCT}$.
In the glassy state it drops again to a value below $0.3 \Omega$.

The drop of the activation volume in the liquid correlates nicely
with the drop of the isotope effect parameter reported for the same
system.\cite{S:01} 

The onset of the increase to the maximum near 
$T_g$ coincides with the onset of a pronounced curvature in the
Arrhenius plot of $D(T)$, Fig.~\ref{fig_D_rho}. At this temperature
and below the isotope effect parameter indicates collective
motion.\cite{S:01} From our previous work \cite{SGO:97} and from the work on
the LJ-system under high pressure \cite{donati:98} we expect a
predominance of chain motion with increasing chain length upon cooling.
At these temperatures diffusion exhibits a pronounced dynamic
heterogeneity over times of several 
$100$~$(m \sigma^2 /\epsilon)^{1/2}$.\cite{CMS:00} The time scale
of this heterogeneity, in turn, correlates with time scale of the
transition from $\beta$- to $\alpha$-relaxation.\cite{kob:95a} 

Our values for the activation energy in the glass are of similar magnitude
as the estimate of Teichler \cite{teichler:01} for the NiZr system.
We find higher activation volumes for less well aged samples. In 
other simulations of metallic glasses
collective jumps of chains of atoms \cite{SOL:93,OS:99,teichler:01}
have been observed, an indication of an inherent collective diffusion
mechanism. 

The experimental values for Ni diffusion in the bulk glass 
Zr$_{46.75}$Ti$_{8.25}$Cu$_{7.5}$Ni$_{10}$Be$_{27.5}$ \cite{knorr:00}
show an increase of the
activation volume below $T=704$~K.
When we plot (open circles in Fig.~\ref{fig_vVFT}) the experimental values in  
LJ units by scaling with the temperature where the break in the 
diffusion constant occurs, i. e. without a free parameter, they
closely follow the LJ result. In particular, they show the
strong increase of $\hat{V}_{\rm act}$ upon cooling towards $T_c^{\rm MCT}$.
This apparent agreement poses, however, some questions.The two temperatures
$T_g$ and $T_c$ nearly coincide  in the
LJ system, under the present quench and aging times. There is experimental
evidence that in  
Zr$_{46.75}$Ti$_{8.25}$Cu$_{7.5}$Ni$_{10}$Be$_{27.5}$ the two temperatures
differ strongly: $T_g = 592$~K and $T_c \approx 875$~K.\cite{meyer:98} 
According to
these numbers the increase of the activation volume would occur at $T_g$
and not at $T_c$ and, therefore, would not be the effect predicted by MCT.
The experimental evidence is not sufficient to clear this point. There
is a large experimental uncertainty on the measured values of the
activation volume, the apparent increase might be an artifact of the fitting. 
Also the MCT critical temperature might be overestimated. Clearly
better experimental values are needed.

As we have seen above, both VFT and MCT expressions predict the onset
of a divergence in $\hat{V}_{\rm act}(T)$. 
The dotted line in Fig.~\ref{fig_vVFT} shows a fit with the
VFT expression, Eq.~\ref{eq_vVFT}. We  assume that the VFT temperatures
are the same for both components, independent of pressure.    
The fit works relatively well.
However, the resulting parameters are somewhat odd. We find:  
${\partial T^{\rm VFT} }/{\partial p}=0.3$ and 
${\partial E^{\rm VFT} }/{\partial p}=0$ for both components and
$\partial \ln{D_0^{\rm VFT}}/{\partial p} = -0.329$ and $-0.2561$ for the
the A and B-components, respectively. That means, according to the fit
there is very little change of the ``activation energy'' and a strong
change of the pre-factor. Allowing for different values of 
${\partial T^{\rm VFT} }/{\partial p}$ the fit is somewhat improved.
Its main shortcoming, the too slow rise, remains.
 
\begin{figure}
\includegraphics[bb=50 100 530 530,totalheight=6cm,keepaspectratio]{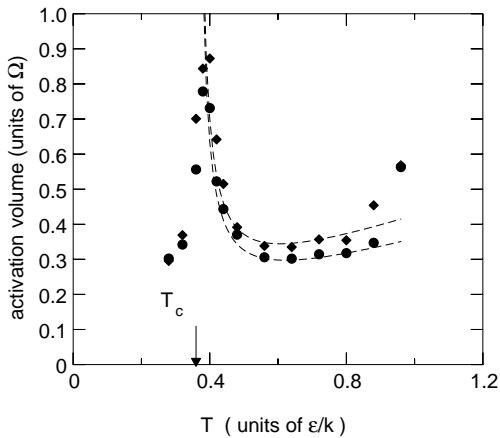}
\caption{Calculated activation volume (majority $A$-atoms, diamonds, 
and minority $B$-atoms, circles) versus temperature.
The dashed line is a fit with the
MCT expression using a common critical temperature for both components.} 
\label{fig_vMCT}
\end{figure}

The alternative fit with the MCT-expression, Eq.~\ref{eq_vMCT}, is displayed
in Fig.~\ref{fig_vMCT}. As above, we assume a common value 
of $T_c^{\rm MCT}$ for both components and all pressures. We omitted in 
the fit the logarithmic term in Eq.~\ref{eq_vMCT}, i.e. we neglected the weak
dependence on a possible pressure variation of $\gamma$. We get an
excellent fit (dashed line) with 
${\partial T_c^{\rm MCT} }/{\partial p} = 0.028$ for both components and 
${\partial \ln{D_0^{\rm MCT}}}/{\partial p} = -0.338$ and $-0.278$ for
A and B atoms,
respectively. The square deviation for this fit is an order 
of magnitude less than the one using the VFT-expression. Contrary
to the VFT case it reproduces  the sharp uprise. We take this as an indication
that the divergence of the activation volume would in deed be
at  $T_c^{\rm MCT}$ and not at $T^{\rm VFT}$. Of course the glass transition
intervenes before the divergence is reached.

Kob and Andersen \cite{kob:95a} report for their constant volume 
simulation a pressure
of $p \approx 2.5$ at their $T_c^{\rm MCT}=0.435$. Extrapolating 
our value using the fitted pressure derivative we find
$T_c^{\rm MCT}(p=2.5) \approx 0.43$ in excellent agreement with 
Kob and Andersen's value.

In conclusion, we find, both in the liquid and the glass, activation volumes
of around 0.3 atomic volumes. 
This correlates with a high collectivity
seen in the isotope effect. In the hot liquid, where diffusion is no longer
by collective motion, the activation volume rises to about 0.6 atomic 
volumes. Cooling towards the critical temperature mode coupling theory
predicts a $1/(T-T_c)$ singularity, cut off by the glass transition. This
is clearly observed in the simulation.
This singular behavior can, e. g. be used to test whether there is
one or several critical temperatures in a multicomponent system.


\end{document}